\documentclass{webofc}
\usepackage[varg]{txfonts}   

\usepackage{amssymb}

\newcommand{\be}{\begin{equation}}
\newcommand{\ee}{\end{equation}}

\begin{document}

\title{Baryons in the plasma: in-medium effects and parity doubling\thanks{Presented at {\em Strangeness in Quark Matter}, Utrecht, the Netherlands, 10-15 July 2017}
}

\author{
	\firstname{Gert} \lastname{Aarts}\inst{1}\fnsep\thanks{Speaker, \email{g.aarts@swan.ac.uk}} 
\and
	\firstname{Chris} \lastname{Allton}\inst{1}
\and
	\firstname{Davide} \lastname{de Boni}\inst{1}
\and
	\firstname{Simon} \lastname{Hands}\inst{1}
\and
	\firstname{Benjamin} \lastname{J\"ager}\inst{2}
\and
	\firstname{Chrisanthi} \lastname{Praki}\inst{1}
\and
	\firstname{Jon-Ivar} \lastname{Skullerud}\inst{3,4}
}

\institute{
	Department of Physics, College of Science, Swansea University, Swansea SA2 8PP, United Kingdom
\and
        Institute for Theoretical Physics, ETH Z\"urich, CH-8093 Z\"urich, Switzerland   
\and
          Department of Theoretical Physics, National University of Ireland Maynooth, County Kildare, Ireland 4
\and
          School of Mathematics, Trinity College Dublin, Dublin 2, Ireland
          }

\abstract{
  We investigate the fate of baryons made out of $u,d$ and $s$ quarks in the hadronic gas and the quark-gluon plasma, using nonperturbative lattice simulations, employing the FASTSUM anisotropic $N_f=2+1$ ensembles.
  In the confined phase a strong temperature dependence is seen in the masses of the negative-parity groundstates, while the positive-parity groundstate masses are approximately temperature independent, within the error. At high temperature parity doubling emerges. A noticeable effect of the heavier $s$ quark is seen. We give a simple description of the medium-dependent masses for the negative-parity states and speculate on the relevance for heavy-ion phenomenology via the hadron resonance gas. 
}

\maketitle

\section{Introduction}

The behaviour of hadrons under the extreme conditions of nonzero temperature and/or density is a longstanding question relevant for e.g.\ heavy-ion collisions and neutrons stars, as well as for the fundamental aspects of chiral symmetry and confinement. Due to the nonperturbative nature of QCD, one has to rely on either models or first-principle studies using the lattice discretisation. 
While mesons at finite temperature have been fairly well studied on the lattice, in various contexts (see e.g.\ Ref.\ \cite{Aarts:2017rrl} for an overview), for baryons this is not the case. In fact, there are only a few lattice studies of baryonic thermal screening \cite{DeTar:1987ar,Pushkina:2004wa} and temporal \cite{Datta:2012fz} masses.  
Here we report on our work on baryons \cite{Aarts:2015mma,Aarts:2017rrl, Aarts:inprep}, using lattice QCD simulations on the FASTSUM anisotropic $N_f=2+1$ ensembles \cite{Aarts:2014cda,Aarts:2014nba}.

We are particularly interested in (the absence of) parity doubling: in the presence of unbroken chiral symmetry, it can be shown that positive- and negative-parity baryonic channels are degenerate, and hence there is parity doubling. When chiral symmetry is (spontaneously) broken, this is no longer the case and indeed, in vacuum the positive-parity groundstate is typically lighter than the negative-parity one. For the nucleon, one finds e.g.\ $m_+\equiv m_N=939$ MeV while $m_-\equiv m_{N^*}=1535$ MeV. Since chiral symmetry is restored around the deconfinement transition,  one expects a degeneracy to emerge. Here we address {\em how} this degeneracy emerges, which is a nonperturbative, dynamical question, not answerable using symmetry considerations alone.

\section{From low to high temperature }

The FASTSUM collaboration uses highly anisotropic lattices, with $a_\tau/a_s \ll 1$, which are constructed specifically for spectral studies of QCD at nonzero temperature. The temperature is varied by changing the number of time slices $N_\tau$,  via the standard relation, $T=1/(a_\tau N_\tau)$, i.e., we use a fixed-scale approach. We have generated several ensembles, four below and four above the deconfinement transition, see Table~\ref{tab-1}. Tuning of the lattice parameters and also the "zero-temperature" ensemble ($N_\tau=128$) were kindly provided by the HadSpec collaboration \cite{Edwards:2008ja}.
The crossover temperature, denoted with $T_c$, is determined via the inflection point of the renormalised Polyakov loop and is found to be $T_c=185(4)$ MeV. This is higher than in nature, due to the light quarks  not having their physical masses ($m_\pi=384(4)$ MeV). The strange quark mass has its physical value, though. For further details, we refer to Refs.~\cite{Aarts:2014cda,Aarts:2014nba}.

\begin{table}[h]
\centering
\caption{Number of time slices $N_\tau$ and the corresponding temperature, in units of $T_c$. Also shown are the number of configurations and sources.  The spatial lattice size is $24^3$, with $a_s= 0.1227(8)$ fm and $a_s/a_\tau=3.5$. } 
\label{tab-1}     
\begin{tabular}{c||cccc | cccc}
\hline
$N_\tau$ & 128 & 40 & 36 & 32 & 28 & 24 & 20 & 16 \\
\hline
$T/T_c$  		& 0.24 & 0.76  & 0.84   & 0.95 & 1.09  & 1.27  & 1.52 & 1.90 \\
$N_{\rm cfg}$ 	& 139  & 501 & 501 & 1000  & 1001 & 1001 & 1000 & 1001\\
$N_{\rm src}$ 	& 16  & 4 & 4 & 2 & 2 & 2 & 2 & 2 \\
\hline
\end{tabular}
\end{table}

\begin{figure}[b]
\centering
\includegraphics[width=0.48\textwidth]{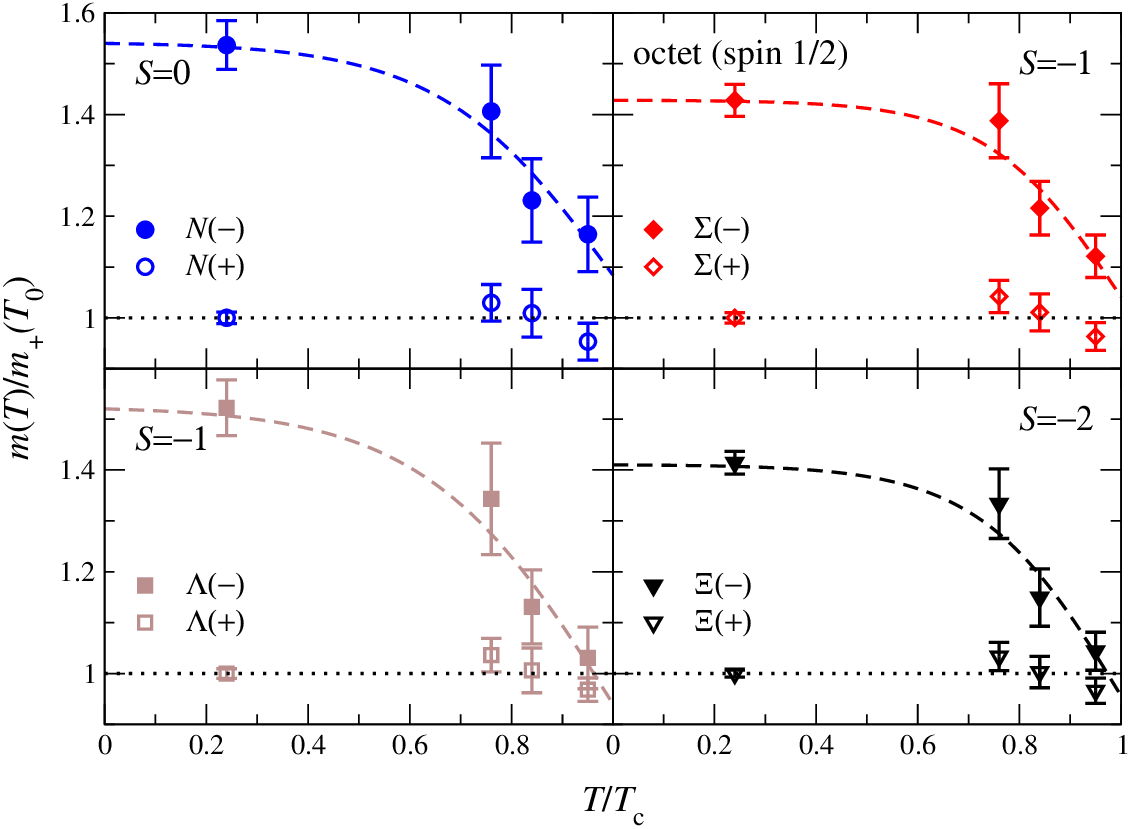}
\includegraphics[width=0.48\textwidth]{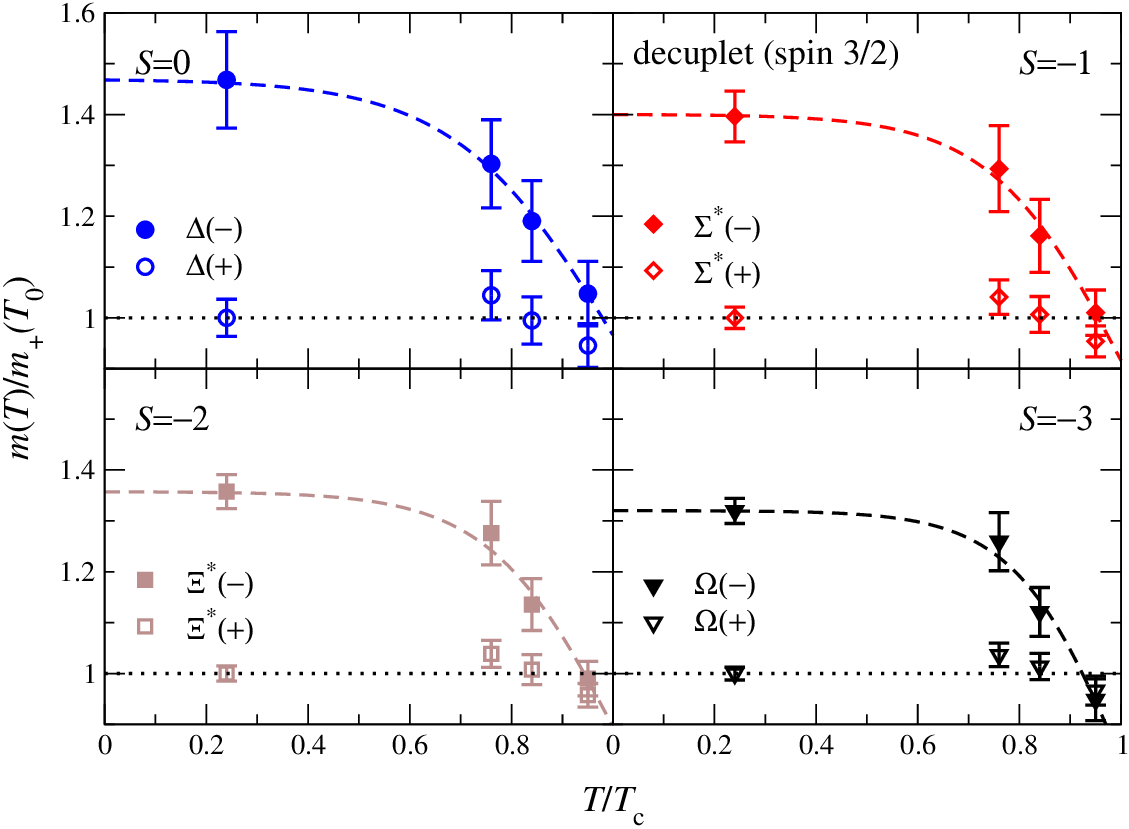}
\caption{Temperature dependence of the groundstate masses, normalised with $m_+$ at the lowest temperature, $m_\pm(T)/m_+(T_0)$, in the hadronic phase, for octet (left) and decuplet (right) baryons. Positive- (negative-) parity masses are indicated with open (closed) symbols.
}
\label{fig:mass}      
\end{figure}

We computed all octet (spin 1/2) and decuplet (spin 3/2) baryon correlators, for both positive and negative parity. Results in the $N, \Delta$ and $\Omega$ channels can be found in Ref.\ \cite{Aarts:2017rrl}; the others ($\Lambda, \Sigma, \Sigma^*, \Xi, \Xi^*$) will appear in Ref.\ \cite{ Aarts:inprep}. In the confined phase, we find that the correlators can be described by combinations of exponentials, allowing us to determine the groundstate masses $m_\pm$ as a function of the temperature. These results are summarised in Fig.\ \ref{fig:mass}, where we show $m_\pm(T)$ in the various channels, normalised with $m_+$ at the lowest temperature.

Several  observations can be made. The positive-parity masses are largely temperature independent. A slight increase and subsequent drop close to $T_c$ can be seen, but it is not significant within current errors. The negative-parity masses on the other hand drop in all channels in a similar way, and become near-degenerate with the corresponding positive-parity mass close to $T_c$. The mass splittings at low temperature are around 500$\sim$600 MeV, in agreement with nature. The dashed lines show a fit using a simple Ansatz, interpolating between $m_-(0)$ and $m_-(T_c)$,
\be
\label{eq:1}
m_-(T) = w(T,\gamma)m_-(0) + \left[1 - w(T,\gamma)\right]m_-(T_c), 
\ee
with the transition function $w(T,\gamma) = \tanh\left[(1 - T/T_c)/\gamma\right]/\tanh(1/\gamma)$, such that 
$w(0,\gamma)=1$ and $w(T_c,\gamma) = 0$. Small (large) $\gamma$ corresponds to a narrow (broad) transition region, which is expected to depend on the masses of the light quarks. 
We have carried out fits in each of the 8 channels and find $0.22\lesssim \gamma\lesssim 0.35$ and $0.85\lesssim m_-(T_c)/m_+(0)\lesssim 1.1$. The largest uncertainty resides in $m_-(T_c)$, since it assumes that the concept of a well-defined groundstate at or close to $T_c$ remains sensible. 

\begin{figure}[t]
\centering
\includegraphics[width=0.453\textwidth]{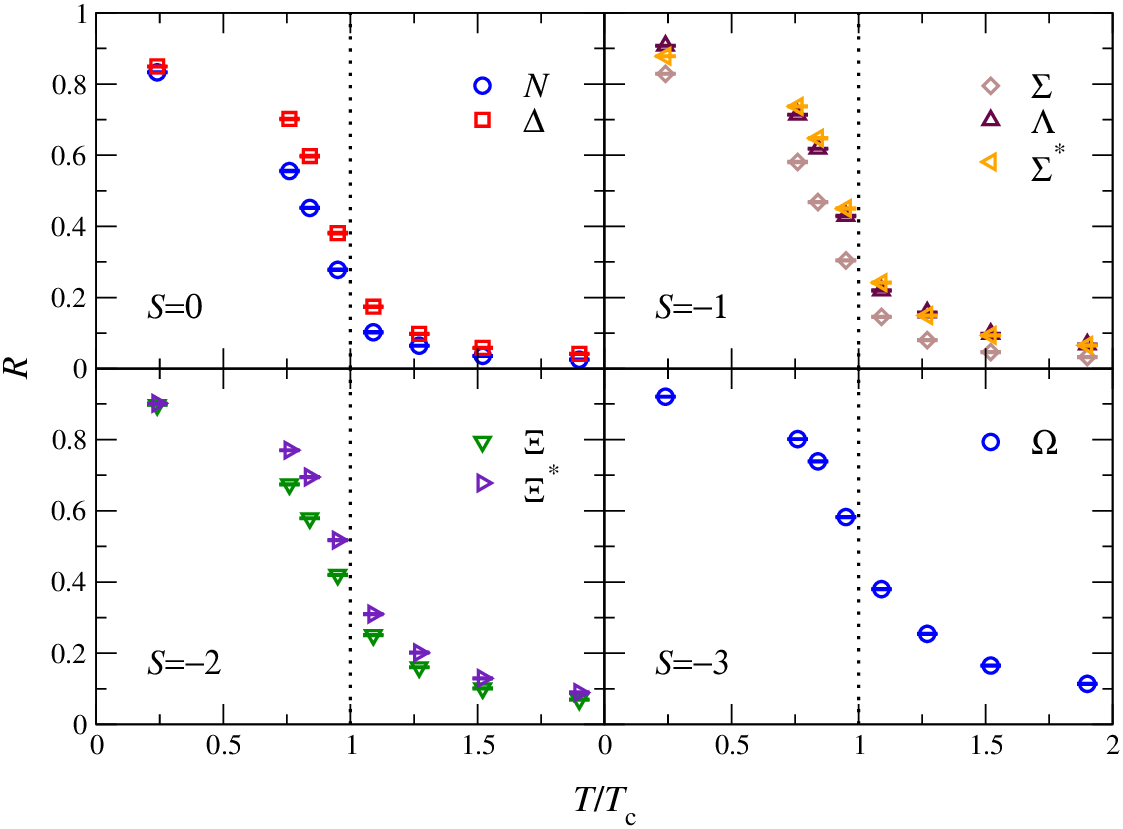}
\includegraphics[width=0.50\textwidth]{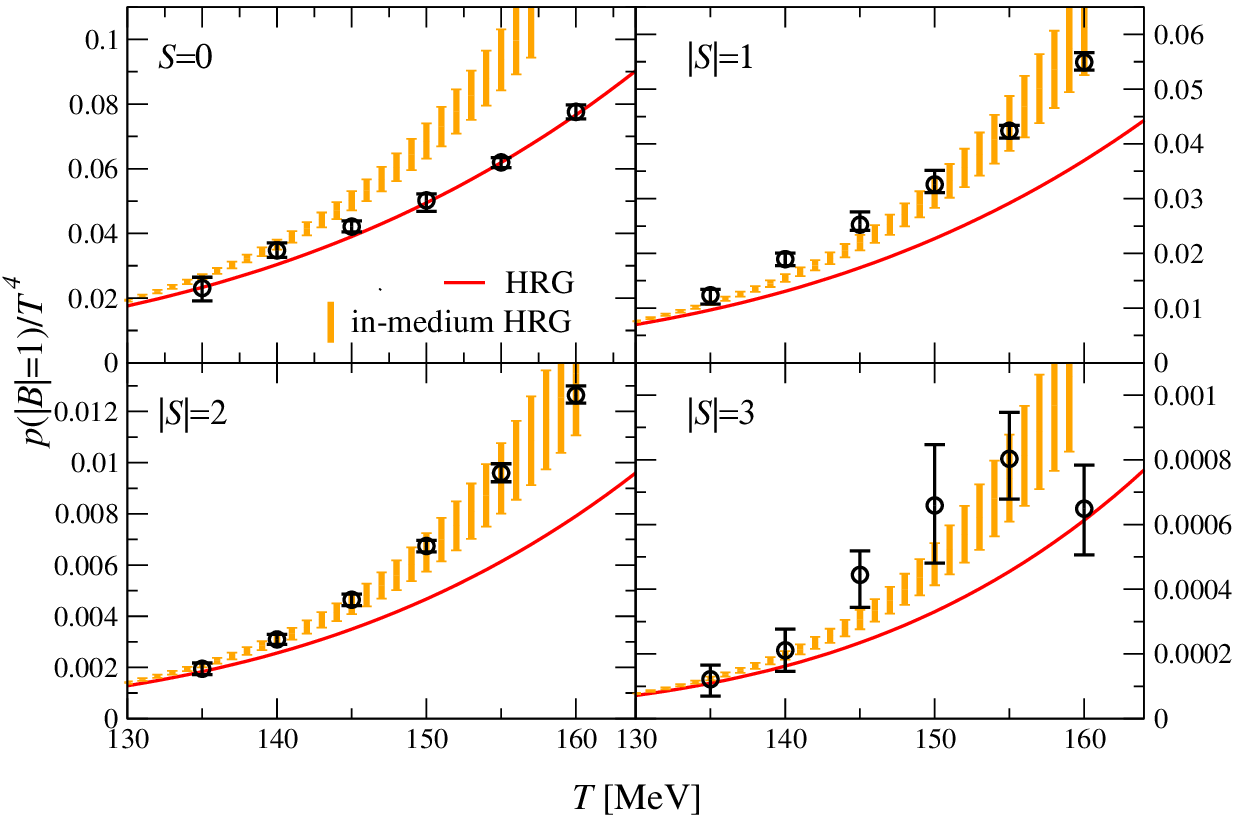}
\caption{Left: temperature dependence of the $R$ ratio, see Eq.~(\ref{eq:2}). Right: contributions to the normalised pressure $p/T^4$ from the sectors with baryon number $|B|=1$ and strangeness $|S|=0,1,2,3$, according to the lattice data of Ref.\ \cite{Alba:2017mqu}, the ideal hadron resonance gas (HRG) and the in-medium HRG, using temperature-dependent negative-parity groundstate masses (\ref{eq:1}), with
 $\gamma = 0.3$, $1<m_-(T_c)/m_+(0) <1.1$ and $T_c=155$ MeV.
}
\label{fig:2}      
\end{figure}

To summarise our findings as $T\to T_c$, we have confirmed that parity doubling indeed emerges, as expected, and established that the degeneracy develops from a reduction of the negative-parity masses with temperature, while the positive-parity masses remain approximately constant. We find this to be similar in all the channels studied.
Effective parity-doublet models \cite{Detar:1988kn} introduce a chirally invariant component to baryon masses to model this type of behaviour and our findings can be used to assess or constrain those models. 
Some recent work along these lines can be found in Refs.\ \cite{Motohiro:2015taa,Mukherjee:2017jzi,Sasaki:2017glk}.

We now turn to the deconfined phase. At our first temperature above $T_c$, $T/T_c=1.09$, no clearly identifiable boundstates appear to be present and e.g.\ exponential fits no longer describe the data  \cite{Aarts:2017rrl}. Hence we focus on the signal for parity doubling directly in the correlators $G_\pm$, using the ratio \cite{Datta:2012fz,Aarts:2015mma}
\be
\label{eq:2}
R(\tau) = \frac{G_+(\tau) - G_-(\tau)}{G_+(\tau) + G_-(\tau)},
\qquad\qquad\qquad
 R = \frac{\sum_n R(\tau_n)/\sigma^2(\tau_n)}{\sum_n 1/\sigma^2(\tau_n)}.
\ee
It should be noted that $G_\pm$  are related via $G_\pm(\tau)=-G_\mp(1/T-\tau)$ and hence $R(1/T - \tau ) = -R(\tau)$. In the case of parity doubling,  $R(\tau)=0$, while in absence of parity doubling and with clearly separated groundstates satisfying $m_- \gg m_+$, $R(\tau)=1$. By summing over the temporal lattice points as in Eq.~(\ref{eq:2}), where $\sigma$ denotes the error, we can construct the quasi-order parameter $R$, which equals 0 in the case of parity doubling and 1 in the case of single states satisfying $m_-\gg m_+$.
 
 The results for $R$ as a function of temperature are shown in Fig.\ \ref{fig:2} (left), organised by strangeness. We note that $R$ indeed changes from 1 to 0 as $T$ increases and that the crossover behaviour coincides with the deconfinement transition.  For baryons with larger strangeness, $R$ is still distinct from 0 at the highest temperatures studied. Since the strange quark mass breaks chiral symmmetry explicitly, its effect is expected to disappear eventually as $m_s/T\to 0$.
 We hence conclude that the interpretation of parity doubling due to chiral symmetry restoration in the baryon sector is indeed valid.

Finally, we consider an application and speculate about the relevance for heavy-ion phenomenology. The distinct medium-dependent masses we found in the confined phase are easily implemented in the hadron resonance gas (HRG). In the usual (ideal) HRG, vacuum masses are used, see e.g.\ Ref.~\cite{Alba:2017mqu} and references therein. In our in-medium modification, we keep the positive-parity masses fixed, but reduce the negative-parity groundstate masses. Concretely, we use the PDG2016 baryon masses classified with 3 and 4 stars, up to 2.5 GeV, and for the in-medium modification (\ref{eq:1}) we take $\gamma = 0.3$, vary $1<m_-(T_c)/m_+(0) <1.1$ (the effect of varying $\gamma$ is negligible) and $T_c=155$ MeV \cite{Borsanyi:2010bp}. The results for the baryonic contributions to the pressure are shown in Fig.\ \ref{fig:2} (right), again organised by strangeness. The ideal and in-medium HRG are compared to the lattice data of Ref.\ \cite{Alba:2017mqu}. While in the $S=0$ sector, the ideal HRG describes the data very well, in the $|S|=1,2,3$ sectors the in-medium HRG leads to an improvement compared to the ideal one. It would hence be interesting to see how this effect compares with other modifications of the HRG that are currently under consideration \cite{Alba:2017mqu,Vovchenko:2016rkn}.

{\bf Acknowledgments}.
We thank Claudia Ratti for providing the lattice data shown in Fig.\ \ref{fig:2} (right).
We are grateful for support from STFC, SNF, ICHEC, the Royal Society, the Wolfson Foundation and COST Action CA15213 THOR. Computing resources were made available by HPC Wales and the STFC-funded DiRAC HPC Facility, which is part of the National E-Infrastructure.

\end{document}